\documentclass[letterpaper]{article} % DO NOT CHANGE THIS
\usepackage{aaai2026}  % DO NOT CHANGE THIS
\usepackage{times}  % DO NOT CHANGE THIS
\usepackage{helvet}  % DO NOT CHANGE THIS
\usepackage{courier}  % DO NOT CHANGE THIS
\usepackage[hyphens]{url}  % DO NOT CHANGE THIS
\usepackage{graphicx} % DO NOT CHANGE THIS
\usepackage{booktabs}
\usepackage{amsmath}
\usepackage{amssymb}
\usepackage{multirow}
\usepackage{natbib}  % DO NOT CHANGE THIS AND DO NOT ADD ANY OPTIONS TO IT
\usepackage{caption} % DO NOT CHANGE THIS AND DO NOT ADD ANY OPTIONS TO IT
\usepackage{algorithm}
\usepackage{algorithmic}

\usepackage{newfloat}
\usepackage{listings}
\DeclareCaptionStyle{ruled}{labelfont=normalfont,labelsep=colon,strut=off} % DO NOT CHANGE THIS
\floatstyle{ruled}
\newfloat{listing}{tb}{lst}{}
\floatname{listing}{Listing}
\title{Hierarchical MoE: Continuous Multimodal Emotion Recognition with Incomplete and Asynchronous Inputs}
\author{
    Yitong Zhu\textsuperscript{\rm 1},
    Lei Han\textsuperscript{\rm 1},
    Guanxuan Jiang\textsuperscript{\rm 1},
    PengYuan Zhou\textsuperscript{\rm 2},
    Yuyang Wang\textsuperscript{\rm 1 *}
}
\affiliations{
    \textsuperscript{\rm 1}the Hong Kong University of Science and Technology (Guangzhou)\\
    \textsuperscript{\rm 2} the Aarhus University\\
    Guangzhou, GuangDong, China\\
    yzhu162@connect.hkust-gz.edu.cn
}

\usepackage{bibentry}
\begin{document}

\maketitle

\begin{abstract}
-Multimodal emotion recognition (MER) is crucial for human-computer interaction, yet real-world challenges like dynamic modality incompleteness and asynchrony severely limit its robustness. Existing methods often assume consistently complete data or lack dynamic adaptability. To address these limitations, we propose a novel Hi-MoE~(Hierarchical Mixture-of-Experts) framework for robust continuous emotion prediction. This framework employs a dual-layer expert structure. A Modality Expert Bank utilizes soft routing to dynamically handle missing modalities and achieve robust information fusion. A subsequent Emotion Expert Bank leverages differential-attention routing to flexibly attend to emotional prototypes, enabling fine-grained emotion representation. Additionally, a cross-modal alignment module explicitly addresses temporal shifts and semantic inconsistencies between modalities. Extensive experiments on benchmark datasets DEAP and DREAMER demonstrate our model's state-of-the-art performance in continuous emotion regression, showcasing exceptional robustness under challenging conditions such as dynamic modality absence and asynchronous sampling. This research significantly advances the development of intelligent emotion systems adaptable to complex real-world environments. The code is uploaded in the Supplementary Material.
\end{abstract}


\section{Introduction}

Multimodal emotion recognition (MER) is critical for applications such as healthcare, social media, and personalized services~\cite{A.V.2023Multimodal, Wang2024Technical}, integrating modalities~\cite{Zheng2019EmotionMeter} to capture complex affective states. However, a fundamental challenge in real-world MER lies in handling dynamic modality incompleteness and asynchrony, where different modalities may be arbitrarily absent or misaligned across various time steps and individual samples. This inherent heterogeneity severely compromises model robustness and generalization, especially for continuous emotion modeling which aims to provide a more nuanced and dynamic representation compared to discrete classification~\cite{Kansizoglou2022Continuous, Romeo2022Multiple}.

Adaptive fusion strategies are essential for handling dynamic and arbitrary modality incompleteness in MER. While methods like early fusion, late fusion, and attention-based integration have been proposed~\cite{Alswaidan2020A, ArunaGladys2023Survey}, they typically assume consistent modality availability. Most approaches focus on enhancing cross-modal interactions or feature alignment to improve performance with complete inputs, with limited emphasis on robustness to arbitrarily missing modalities (e.g., varying absences across time steps or samples). Simple techniques, such as zero-imputation or modality dropout, often yield suboptimal predictions and significant performance degradation. However, designing sophisticated fusion mechanisms for robust continuous emotion modeling under dynamic modality absence remains underexplored. 

Recent studies have explored Mixture-of-Experts (MoE) architectures in multimodal learning to address incomplete modality availability~\cite{Wu2023Omni-SMoLA:, Xu2024Leveraging}. By conditionally activating expert modules, MoE offers structural flexibility and has shown initial promise in handling modality uncertainty. However, existing MoE implementations~\cite{Zhou2022Mixture-of-Experts, Dai2022StableMoE} often rely on hard routing or single-path activation, which restricts inter-modal collaboration and flexible integration of partial information. Furthermore, their application has predominantly focused on shallow architectures and discrete classification, consequently leaving their efficacy for continuous emotion regression largely uninvestigated within the MoE paradigm. A critical limitation is the common assumption of strict temporal alignment; signal dynamics differences between different modalities~\cite{Binias2020Prediction} often cause semantic misalignment. Collectively, these limitations reveal a significant gap: current MoE models primarily address modality-level variations, with limited exploration of their potential to model the complex, continuous structure of emotions for robust affective state prediction.

Motivated by these limitations, we propose a novel frame work, Hi-MoE (hierarchical MoE), for multi-dimensional continuous emotion regression. Our framework features two key components: a Modality Expert Bank that uses soft routing for robust inference under missing modalities, and an Emotion Expert Bank with attention-based routing to select informative emotional prototypes within the fusion space. This dual-level architecture uniquely balances flexible modality adaptation with selective and semantically meaningful representation learning. Furthermore, it incorporates a cross-modal alignment module to explicitly handle temporal shifts and semantic inconsistencies, enhancing the model’s capacity to capture complementary cues across a wide range of modalities. We demonstrate the effectiveness of Hi-MoE on real-world datasets, including the DEAP and the DREAMER, which involves several key modalities including behavioral and physiological modalities. The results confirm the robustness and generalization of Hi-MoE in diverse scenarios of incomplete and asynchronous inputs. The main contributions of this study are summarized as follows:
 % Overall, our approach enables highly adaptable, scalable, and generalizable continuous emotion modeling for real-world applications. 
\begin{itemize}
\item We propose a Hierarchical MoE framework that effectively incorporates arbitrary modality combinations and robustly addresses the missing modality scenario, with soft-routing for modality handling and differential-attention routing for emotion prototype modeling.
\item We incorporate a dedicated cross-modal alignment module that explicitly handles temporal shifts and semantic inconsistencies between different modalities.
\item Experiments on DEAP and DREAMER showcase the
consistent and robust performance of Hi-MoE in handling diverse modality combinations.
\end{itemize}

\section{Related Work}
\subsection{Multimodal Emotion Recognition Methods}
Multimodal emotion recognition (MER) integrates complementary cues from modalities like physiological signals and behavioral cues to enhance accuracy via cross-modal fusion. Early works, such as Poria et al.~\shortcite{Poria2015104Towards} and Williams et al.~\shortcite{williams2018recognizing}, employed early and late fusion for unified emotion representations. Subsequent studies introduced LSTM-based models for intra-modality temporal modeling~\cite{Huang2017Continuous, Su2020AnImproved} and self-attention mechanisms for long-range dependencies~\cite{Sun2020MultimodalContinuous, Su2021MultimodalEmotion}. Hybrid approaches are usee to balanced early and late fusion benefits~\cite{Nemati2019AHL}.

Despite these advances, existing methods often fail to capture fine-grained inter-modality interactions or rely on modality-specific pipelines, limiting their flexibility in complex scenarios. Intermediate fusion strategies, such as Self-MM~\cite{YuXuYuanWu2021Learning} and Han et al.~\shortcite{han2021improving}, integrate modalities within network layers to improve cross-modal modeling. Transformer-based architectures further enhance dynamic inter-modal dependencies~\cite{Zadeh2018Multiattention, mittal2020m3er}, with some addressing cross-modal asynchrony~\cite{lv2021progressive}. However, most approaches assume complete and synchronized multimodal inputs, rendering them less effective in real-world settings with missing or degraded data. Our work aims to fill this critical gap by developing robust and adaptive fusion mechanisms that adapt to varying modality availability and explicitly manage asynchronous responses or semantic disparities.

\subsection{Mixture of Experts for Multimodal Learning}
The Mixture-of-Experts (MoE) framework, first proposed by Jacobs et al.~\shortcite{jacobs1991adaptive} and Jordan et al.~\shortcite{jordan1994hierarchical}, employs a gating mechanism to route inputs to specialized expert networks, enabling dynamic resource allocation and efficient scaling for high-capacity tasks~\cite{fedus2022switch}. In deep learning, Shazeer et al.~\shortcite{shazeer2017outrageously} advanced MoE with sparsely-gated Transformers, followed by developments like GShard for large-scale translation~\cite{lepikhin2020gshard}, Switch Transformer for simplified routing~\cite{fedus2022switch}, and MoEfication for transforming dense models~\cite{riquelme2021scaling}. Stable training strategies were also introduced~\cite{roller2020recipes}. MoE has been adapted for multi-task learning, with frameworks like VmoE~\cite{riquelme2021scaling} and TaskMoE~\cite{ye2023taskexpert} using task-aware gating to select relevant experts.

Recently, MoE has been applied to multimodal learning to address cross-modal heterogeneity. Cheng et al.~\shortcite{cheng2024mixtures} developed a dynamic fusion MoE for audio-visual speech recognition, while Huai et al.~\shortcite{huai2025cl} proposed a dual-router MoE for continual learning across modalities. In emotion recognition, MoE has been used to align modality-specific and invariant features~\cite{Gao2024Enhanced, fang2025emoe}. However, these approaches often neglect temporal dynamics and modality-specific semantics, relying on shallow structures and rigid routing, which limits their expressiveness and robustness to missing or asynchronous inputs. To address these limitations, the proposed hierarchical MoE framework employs adaptive routing and conditional structure selection to effectively handle cross-modal heterogeneity and temporal dynamics for continuous emotion prediction.

\section{Methodology}
\begin{figure*}[htbp]
    \centering
    \includegraphics[width=1.0\linewidth]{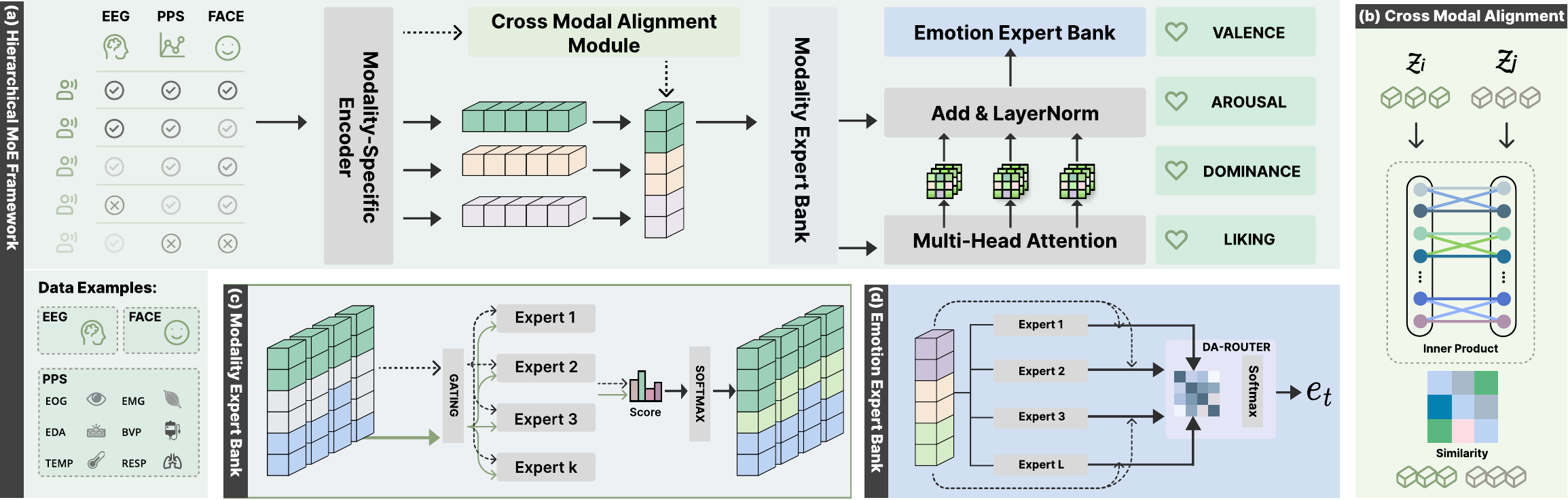}
    \caption{Overall Architecture of the Proposed Hierarchical MoE Framework. (a) The hierarchical MoE framework processes raw multimodal signals. The completeness of each modality at a given time step is visually represented by the shading intensity of its corresponding feature block after modality-specific encoding (darker indicates presence, lighter/faded indicates absence). (b) The Cross-Modal Alignment module explicitly mitigates temporal and semantic discrepancies between encoded modality features during training. (c) The Modality Expert Bank showcases its per-modality soft gating mechanism and expert processing for robust information fusion. (d) The Emotion Expert Bank illustrates how the Differential Attention Router dynamically attends to specialized emotion experts to model fine-grained emotional prototypes.}
    \label{fig:framework_hier_moe}
\end{figure*}

This section presents the Hierarchical Mixture-of-Experts (Hi-MoE) framework for robust continuous emotion recognition with incomplete and asynchronous multimodal inputs. As shown in Fig.~\ref{fig:framework_hier_moe}, the architecture features a dual-layer expert structure: a Modality Expert Bank to adaptively process diverse input modalities and handle missing data, and an Emotion Expert Bank to capture fine-grained emotional prototypes. A cross-modal alignment module is integrated to address temporal and semantic inconsistencies across modalities. The framework employs adaptive routing and conditional structure selection to enhance robustness and expressiveness. The following subsections detail the design of it. While exemplified using eight modalities shown in the figure, the proposed model is generalizable to various modality combinations.

\subsection{Modality-Specific Encoder}
Considering the structural and semantic differences across modalities, we introduce modality-specific encoders to project inputs into a unified feature space. To enhance sensitivity to short-term emotional dynamics, a sliding window is applied to all raw modalities, increasing temporal density and promoting local variation awareness.

We employ various encoder architectures, each specifically tailored to its respective modality's characteristics, to effectively capture its unique patterns. To preserve modality-specific semantics, parameters are not shared across encoders. The detailed architectures of these encoders are provided in the Appendix. The resulting features $\{ \mathbf{Z}_m \}$ are then temporally aligned and serve as input to the subsequent expert routing and fusion modules.

\subsection{Cross-Modal Alignment Module}
Direct temporal fusion of raw or fused multimodal representations often overlooks response delays inherent in MER, leading to semantic misalignment and unreliable emotion trajectory modeling. To address this, we introduce a contrastive learning-based alignment module to mitigate temporal lags and semantic discrepancies in the latent space.

For each pair of modality representations $[\mathbf{Z}_i, \mathbf{Z}_j] \in \mathbb{R}^{T \times d}$, where $T$ is the number of time steps and $d$ is the feature dimension, derived from the modality-specific encoders, we apply $\ell_2$ normalization and compute a similarity matrix over the concatenated batch:

\begin{equation}
S_{i,j} = \frac{\mathbf{Z}_i \cdot \mathbf{Z}_j}{\tau}, \quad \mathbf{Z} = [\mathbf{Z}_i; \mathbf{Z}_j] \in \mathbb{R}^{2T \times D}
\label{eq:similarity_matrix}
\end{equation}
where $\tau$ is a temperature hyperparameter, and diagonal entries of $S$ are masked to avoid trivial self-similarity. The NT-Xent loss is employed to align semantically similar representations across modalities despite temporal misalignments. The training details provided in the next subsection.

\subsection{Hierarchical MoE Structure}

\subsubsection{Modality Expert Bank}
To address modality heterogeneity, missing data, and varying signal quality, we propose a Modality Expert Bank as the first layer of the hierarchical framework. This component adaptively processes diverse input modalities while handling absence or asynchrony. At each time step $t$, the input from all potential modalities is denoted as $
\mathbf{X}_t = \{ \mathbf{x}_t^{(1)}, \mathbf{x}_t^{(2)}, \dots, \mathbf{x}_t^{(M)} \}
$, where $\mathbf{x}_{t}$ denotes the feature vector for modality $m$, and $M$ is the total number of modalities. For missing modalities, $m$ is set to a zero vector of appropriate dimension, ensuring a consistent input structure while indicating the absence of data.

Each modality $m$ is associated with an expert bank $\mathcal{E}^{(m)} = \{ E^{(m)}_1, \dots, E^{(m)}_K \}$, comprising $K$ learnable expert networks. A modality-specific soft gating mechanism computes routing weights as :
\begin{equation}
\alpha_t^{(m)} = \text{Softmax}( W_g^{(m)} \mathbf{x}_t^{(m)} + b_g^{(m)} ) \in \mathbb{R}^K
\end{equation}
where $W_g^{(m)}$ and $b_g^{(m)}$ are learnable parameters. For present modalities, $\alpha_t^{(m)}$ dynamically routes $\mathbf{x}_t^{(m)}$ to the corresponding experts. For missing modalities $\mathbf{x}_t^{(m)} = \mathbf{0}$, the gating mechanism relies primarily on the bias term $b_g^{(m)}$ to produce default weights, signaling the absence of valid input. The modality-specific output is computed as a weighted sum of expert responses:
\begin{equation}
    \mathbf{z}_t^{(m)} = \sum_{k=1}^K \alpha_t^{(m)}[k] \cdot E_k^{(m)}( \mathbf{x}_t^{(m)} )
\end{equation}
where $\alpha_t^{(m)}[k]$ denotes the $k$-th element of the gating weights. This adaptive routing mechanism enables robust processing regardless of modality availability. The modality-specific representations are then fused into a unified embedding:
\begin{equation}
    \label{eq: fusion}
    \mathbf{z}_t = \textit{Fuse}\left( \mathbf{z}_t^{(1)}, \dots, \mathbf{z}_t^{(M)} \right)
\end{equation}
where $\textit{Fuse}(\cdot)$ is an attention-based aggregation mechanism that integrates modality-specific features. The fused representation $\mathbf{z}_t$ serves as input to the subsequent Emotion Expert Bank, detailed in the next subsection.

\subsubsection{Emotion Expert Bank}
To model fine-grained emotional prototypes and temporal dynamics in continuous emotion recognition, we propose an Emotion Expert Bank as the second layer of the hierarchical MoE framework. This component processes the fused multimodal representation $\mathbf{z}_t$ from the Modality Expert Bank (Eq.~\ref{eq: fusion}) to generate emotion-specific features. Unlike traditional MoE frameworks that employ static or modality-agnostic routing, we introduce a Different Attention Router (DA-router) that leverages attention-based mechanisms to dynamically allocate $\mathbf{z}_t$ to specialized emotion experts based on temporal and contextual variations.

The Emotion Expert Bank consists of $L$ learnable expert networks $\mathcal{F} = \{ F_1^{\text{emo}}, \dots, F_L^{\text{emo}} \}$, each designed to capture distinct emotional prototypes. At time step $t$, the DA-router computes attention-based routing weights as:

\begin{equation}
\beta_t[l] = \frac{\exp\left( \phi\left(\mathbf{z}_t, F_l^{\text{emo}}(\mathbf{z}_t) \right) \right)}{\sum_{j=1}^L \exp\left( \phi\left(\mathbf{z}_t, F_j^{\text{emo}}(\mathbf{z}_t) \right) \right)}, \quad l = 1, \dots, L,
\label{eq:da_router}
\end{equation}
where $\phi(\cdot, \cdot)$ is a similarity function that measures the relevance between the input $\mathbf{z}_t$ and the output of the $l$-th emotion expert $F_l^{\text{emo}}(\mathbf{z}_t)$. This differential attention mechanism enables the DA-router to prioritize experts based on their alignment with the emotional context. The output of this module is computed as:

\begin{equation}
\mathbf{e}_t = \sum_{l=1}^L \beta_t[l] \cdot F_l^{\text{emo}}(\mathbf{z}_t),
\label{eq:emotion_output}
\end{equation}
where $\beta_t[l]$ represents the weight for the $l$-th expert. The DA-router’s attention-based design enhances adaptability to complex emotional patterns under diverse input conditions. The resulting representation $e_{t}$ serves as the final emotion-specific feature for downstream prediction heads.

\subsection{Dimension-Specific Prediction and Loss}
To model diverse emotion dimensions without task interference, we use independent output heads for each dimension. The model supports two prediction formats: (1) binary classification for high/low levels and (2) ordinal regression on a 1–9 Likert scale for fine-grained intensity estimation.

For cross-modal alignment, positive pairs are defined as representations of the same sample across different modalities. The Normalized Temperature-scaled Cross Entropy (NT-Xent) loss~\cite{agren2022ntxentlossupperbound} is formulated as:

\begin{equation}
\mathcal{L}_{\text{NT-Xent}} = -\frac{1}{2B} \sum_{i=1}^{2B} \log \frac{\exp(S_{i,y_i})}{\sum_{j=1}^{2B} \textit{1}_{[j \neq i]} \exp(S_{i,j})}
\end{equation}
where $S_{i,j}$ is the similarity score from Eq.~\ref{eq:similarity_matrix}, and $y_i$ denotes the index of the positive sample paired with $i$.

For emotion prediction, we use cross-entropy loss for binary classification and mean squared error (MSE) for ordinal regression. A label mask excludes missing labels from gradient updates, enhancing training stability. When the cross-modal alignment module is enabled, the alignment loss is jointly optimized with the emotion prediction loss. The total training loss is:

\begin{equation}
\mathcal{L} = \mathcal{L}_{emo} + \lambda \cdot \mathcal{L}_\text{NT-Xent}
\end{equation}
where $\mathcal{L}{\textit{emo}}$ represents the emotion prediction loss, $\mathcal{L}_{\text{NT-Xent}}$ is the contrastive alignment loss, and $\lambda$ is a tunable coefficient balancing the two objectives. Training details are discussed in the following section.

\section{Experiments}

\subsection{Experimental setup}
\subsubsection{Dataset}
To evaluate the performance of our proposed model on emotion recognition, we conduct experiments on two widely used multimodal emotion datasets: DEAP~\cite{Koelstra2012DEAP} and DREAMER~\cite{Katsigiannis2018DREAMER}. Both datasets provide synchronized signals with continuous emotional annotations, offering a suitable benchmark for evaluating fine-grained affective modeling, as summarized in Table~\ref{tab:dataset_comparison}. For brevity, we refer to the four affective dimensions, Valence(V), Arousal(A), Dominance(D), and Liking(L), in the following sections. The more comprehensive description of the two datasets and their pre-processing is shown in the Appendix. 

\begin{table}[ht]
\centering
\caption{Comparison of DEAP and DREAMER datasets}
\label{tab:dataset_comparison}
\begin{tabular}{ccc}
\toprule
 & \textbf{DEAP} & \textbf{DREAMER} \\
\midrule
Number & 32 & 23 \\
Video & Musical videos & Movie Clips \\
Trials & 40 & 18 \\
Physiological & EEG, PPS & 2 \\
Facial & Yes & No \\
Annotation & V-A-D-L & V-A-D \\
\bottomrule
\end{tabular}
\end{table}

\subsubsection{Metrics}
We evaluate model performance using metrics suitable for both continuous regression and binary classification tasks. For continuous prediction, we report the Mean Absolute Error (MAE), the Concordance Correlation Coefficient (CCC) and Pearson Correlation Coefficient(PCC). Besides, we report Accuracy (ACC)
and F1-score for binary emotion classification. All metrics are reported for each emotion dimension where applicable.

\subsubsection{Experimental Settings}
All experiments were conducted on six NVIDIA RTX 4090 GPUs using PyTorch 1.13.1. We evaluated the model under both subject-dependent and subject-independent protocols, using 10-fold cross-validation and Leave-One-Subject-Out (LOSO) validation, respectively. Results are reported as the mean ± standard deviation across folds or subjects. Models were trained for 75 epochs with Adam optimizer, cosine annealing, batch size 32 per GPU, and early stopping to prevent overfitting.

\subsubsection{Baseline Methods}

We compare Hi-MoE with a set of representative and state-of-the-art multimodal fusion baselines commonly used in emotion recognition tasks. These include late fusion~\cite{Tang2017Multimodal},  hybrid fusion Bi-LSTM~\cite{zhao2021expression}, attention-based TACOFormer~\cite{li2023tacoformer}, and transformer fusion IANet~\cite{LI2024IANet}. All of which have demonstrated strong performance on benchmark datasets mentioned before. All baselines are implemented under consistent preprocessing, input formats, and training protocols to ensure fair comparison. We also adapt them to support continuous emotion regression by replacing the classification head with regression layers and using the same evaluation metrics introduced above. And * in the tables indicates that the results of this model were reproduced by ourselves, while others are from the literature. 

\subsection{Subject-Dependent Experiments}

We first evaluate the proposed method in a subject-dependent setting using 10-fold cross-validation, with each subject’s trials evenly split across folds. The following subsections present results on emotion regression, binary classification, robustness to missing modalities, and ablation studies. Unless noted otherwise, performance is reported using the previously introduced metrics.

\subsubsection{Performance on Complete Multimodal Data}
To establish our model's baseline ability to capture fine-grained emotional variations using complete multimodal inputs and to demonstrate its generalizability across diverse datasets, we evaluate the proposed Hierarchical MoE framework on continuous emotion regression across four affective dimensions. We conduct experiments on multiple benchmark datasets, including DEAP and DREAMER, which are shown in Tab.~\ref{tab:main_experiments}. We construct a fair and consistent comparison with representative baselines by training all methods under the same modality configurations and data preprocessing procedures for each respective dataset.

\begin{table*}[htbp]
\centering
\caption{Overall average performance comparison of different fusion methods on emotion regression. (CCC, PCC$\uparrow$ MAE$\downarrow$)}
\label{tab:main_experiments}
\begin{tabular}{lccccccc} % Columns: Method, Modalities, DEAP (CCC, PCC, MAE), DREAMER (CCC, PCC, MAE)
\toprule
\multirow{3}{*}{\textbf{Method}} & \multirow{3}{*}{\textbf{Modalities}} & \multicolumn{3}{c}{\textbf{DEAP}} & \multicolumn{3}{c}{\textbf{DREAMER}} \\
\cmidrule(lr){3-5} \cmidrule(lr){6-8}
 & & CCC & PCC & MAE & CCC & PCC & MAE \\
\midrule
Late Fusion~\shortcite{Tang2017Multimodal} & EEG, PPS & 0.8541 & 0.8827 & 0.755 & 0.370 & 0.400 & 0.855 \\
TACOFormer*~\shortcite{li2023tacoformer} &  EEG, PPS & 0.9144 & 0.9379 & 0.721 & 0.9423 & 0.9498 & 0.696\\ 
\textbf{Ours} & \textbf{EEG, PPS} & \textbf{0.9522} & \textbf{0.9679} & \textbf{0.655} & \textbf{0.9571} & \textbf{0.9623} & \textbf{0.612} \\
\midrule
Bi-LSTM*~\shortcite{zhao2021expression} & EEG, PPS, Face & 0.8712 & 0.8899 & 0.4626 & --- & --- & --- \\
\textbf{Ours} & \textbf{EEG, PPS, Face} & \textbf{0.9698} & \textbf{0.9790} & \textbf{0.321} & --- & --- & --- \\
\midrule
IANet*~\shortcite{LI2024IANet} &  EEG, EOG, EMG, EDA & 0.9002 & 0.9207 & 0.737 & 0.9114 & 0.9288 & 0.729 \\
\textbf{Ours} &  \textbf{EEG, EOG, EMG, EDA} & \textbf{0.9131} & \textbf{0.9218} & \textbf{0.711} & \textbf{0.9310} & \textbf{0.9402} & \textbf{0.702} \\
\bottomrule
\end{tabular}
\end{table*}

To further illustrate the model's ability to capture fine-grained emotional changes, we visualize the predicted trajectories over time in comparison with the ground truth. Fig.~\ref{fig:V-A-trajectories-over-time} shows the V-A dimension (others are in the Appendix). Our model closely tracks the emotional trend while producing smoother and more temporally consistent predictions than baseline methods.

\begin{figure}[htbp]
\centering
\includegraphics[width=0.99\linewidth]{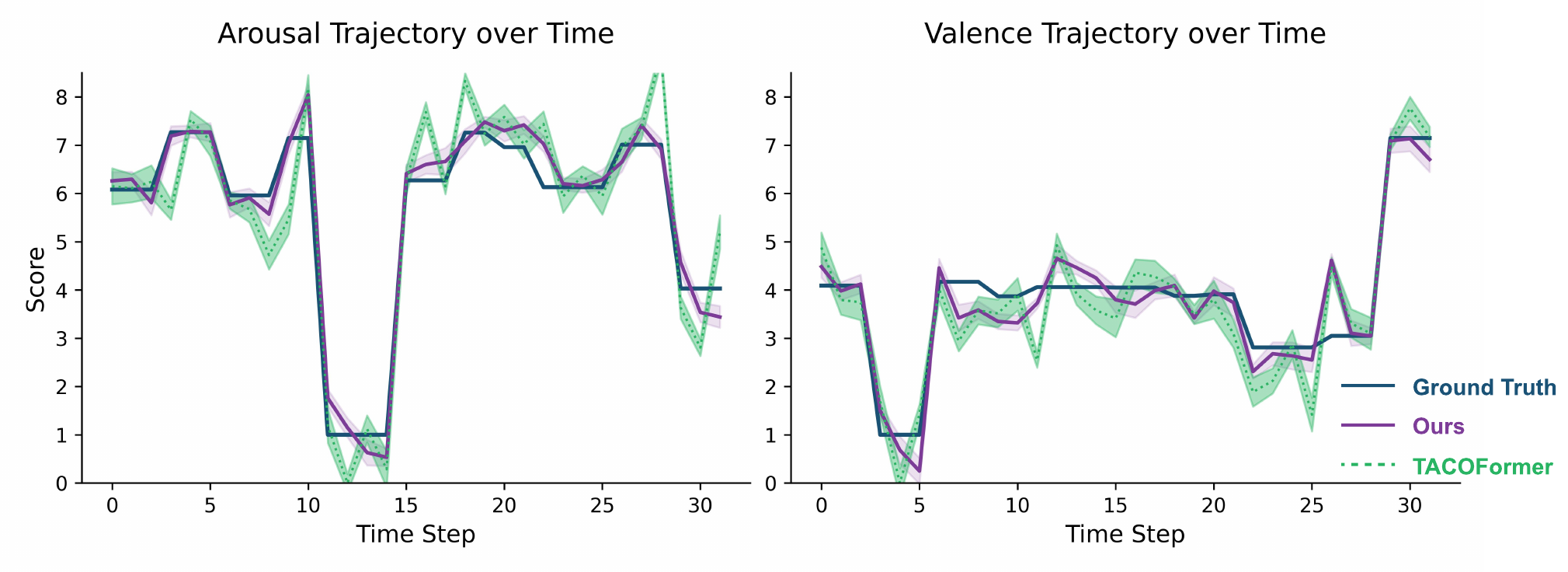}
\caption{Predicted trajectories compared with ground truth.}
\label{fig:V-A-trajectories-over-time}
\end{figure}

\subsubsection{Robustness under Modality Missing}

In real-world multimodal scenarios, input modalities are often partially unavailable due to sensor failures, occlusion, or transmission noise. To evaluate Hi-MoE's robustness under such conditions, we conduct controlled experiments with varying modality missing rates during inference. We focus on the all-modality setting, as no missing input, which represents a typical multimodal configuration with complementary physiological signals. Additional results with other combinations for other combinations are provided in the Appendix. Specifically, we simulate incomplete inputs by randomly masking arbitrary combinations of modalities for each sample in both training and evaluation. The missing rate $r$ ranges from 0.00 to 0.40 in increments of 0.05. Each configuration is repeated with five random seeds, and average performance is reported to reduce the influence of stochastic variation.

\begin{figure}[!h]
    \centering
    \includegraphics[width=1.0\linewidth]{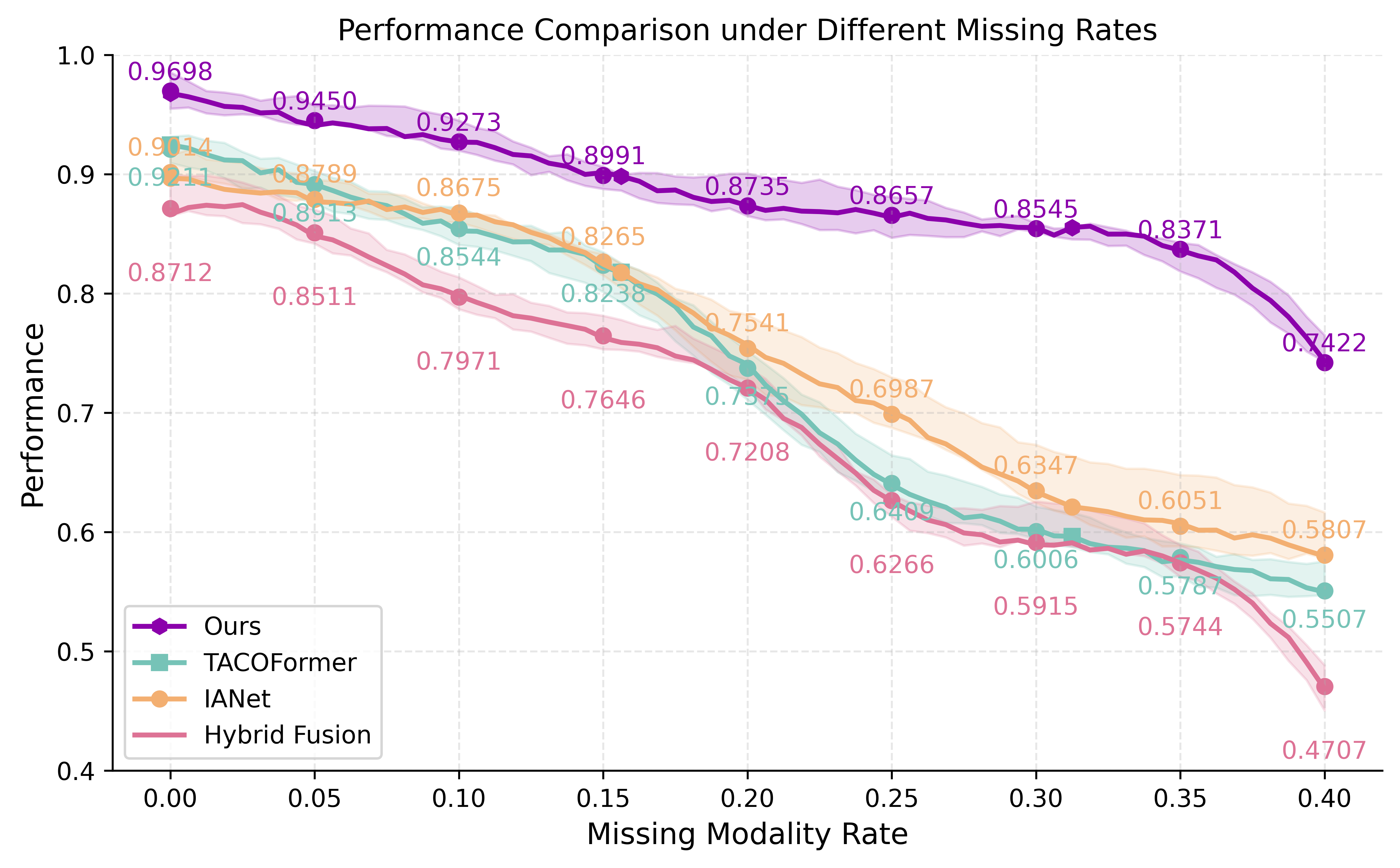}
    \caption{Performance comparison under varying modality missing rates. Shaded regions indicate standard deviation across different random seeds.}
    \label{fig:performance_under_different_rates}
\end{figure}

Tab.~\ref{tab:modality_missing} presents the precise CCC values across different modality missing rates for Hi-MoE and various fusion-based baselines. Complementing this, Fig.~\ref{fig:performance_under_different_rates} visually illustrates the performance trends. As expected, the performance of all models degrades as the missing rate increases. However, the proposed model consistently exhibits significantly slower performance degradation compared to the baselines mentioned before. At a 0.00 missing rate, it achieves a CCC of 0.9698, and even with a high 35\% missing input, it maintains a competitive CCC of 0.8371, which is substantially higher than Hybrid Fusion (0.5744), IANet (0.6051), and TACOFormer (0.5744) at the same missing rate. This remarkable resilience to modality loss, even up to 40\% missing inputs (CCC of 0.7422), strongly highlights the effectiveness of the Modality Expert Bank in dynamically re-routing computation and preserving task performance under incomplete signals. These results underscore the framework's superior robustness, making it highly suitable for real-world applications where data integrity cannot always be guaranteed.

These results demonstrate that each component of the proposed framework contributes meaningfully to its overall performance and highlight the importance of hierarchical structure, adaptive expert routing, and temporal alignment in continuous emotion recognition.
\begin{table*}[htbp]
\centering
\caption{CCC under different modality missing rates on the DEAP dataset (CCC $\uparrow$).}
\label{tab:modality_missing}
\begin{tabular}{lccccccccccc}
\toprule
Missing Rate ($r$) & 0.00 & 0.05 & 0.10 & 0.15 & 0.20 & 0.25 & 0.30 & 0.35 & 0.40 \\
\midrule
Bi-LSTM*~\shortcite{zhao2021expression}    & 0.8712 & 0.8511 & 0.7971 & 0.7646 & 0.7208 & 0.6266 & 0.5915 & 0.5744 & 0.4707 \\
IANet*~\shortcite{LI2024IANet}     & 0.9014 & 0.8789 & 0.8675 & 0.8265 & 0.7541 & 0.6987 & 0.6347 & 0.6051 & 0.5807 \\
TACOFormer*~\shortcite{li2023tacoformer}& 0.9211 & 0.8913 & 0.8544 & 0.8238 & 0.7375 & 0.6409 & 0.6006 & 0.5787 & 0.5507 \\
\textbf{Ours (Full)} & \textbf{0.9698} & \textbf{0.9450} & \textbf{0.9273} & \textbf{0.8991} & \textbf{0.8735} & \textbf{0.8657} & \textbf{0.8545} & \textbf{0.8371} & \textbf{0.7422} \\
\bottomrule
\end{tabular}
\end{table*}

\subsubsection{Discrete Classification with Adaptive Modality Handling}
While the primary goal is continuous emotion regression, we also evaluate the framework's generalization ability on standard binary classification tasks. To adapt our model, we replace the regression head with a binary classification head, trained using binary cross-entropy loss, with no other architectural changes.

To ensure a fair and comprehensive comparison with existing baselines, which often report their best performance on binary classification using specific, and sometimes varying, modality subsets, we conduct evaluations under several common modality configurations with the help of the inherent flexibility in handling arbitrary modality combinations of our model. As shown in Table~\ref{tab:binary_results}, Hi-MoE consistently achieves higher accuracy and F1-score on both V and A classification on the DEAP dataset with the input of the EEG and PPS. And the results of other relevant subsets are shown in the Appendix. These results collectively demonstrate that the Hierarchical MoE retains strong discriminative capability and exhibits remarkable cross-task flexibility and robustness, even when operating with constrained or partial input modality subsets.

\begin{table}[!h]
\centering
\caption{Performance on binary classification (V-A).}
\label{tab:binary_results}
\begin{tabular}{lcccc}
\toprule
\multirow{2}{*}{\textbf{Method}} & \multicolumn{2}{c}{\textbf{Valence}} & \multicolumn{2}{c}{\textbf{Arousal}} \\
\cmidrule(lr){2-3} \cmidrule(lr){4-5}
 & ACC & F1 & ACC & F1 \\
\midrule
Late Fusion~\shortcite{Tang2017Multimodal}             & 83.23 & - & 88.82 & - \\
TACOFormer~\shortcite{li2023tacoformer}  & 91.59 & - & 92.02 & - \\
MFST-RNN~\shortcite{LI2023MFSTRNN}   & 94.99 & 95.40 & 95.89 & 96.09 \\
\textbf{Ours}  & \textbf{98.15} & \textbf{98.77} & \textbf{98.63} & \textbf{98.91}\\
\bottomrule
\end{tabular}
\end{table}

\subsection{Subject-Independent Evaluation}
To evaluate the generalization capability of Hi-MoE across unseen individuals, we conduct subject-independent experiments using a LOSO cross-validation protocol. In this setup, for each fold, data from one subject is held out for testing while the model is trained on all remaining participants. This rigorous protocol simulates real-world scenarios where emotion recognition systems must operate on users not seen during training. Tab.~\ref{tab:subject_independent} reported the average metrics across all subjects, the proposed framework also achieves strong generalization performance, consistently outperforming baseline fusion methods, which demonstrate the robustness of the hierarchical soft expert design in modeling individual-invariant affective patterns.
\begin{table}[htbp]
\centering
\caption{LOSO on DEAP dataset. Metrics: CCC (↑).}
\label{tab:subject_independent}
\begin{tabular}{lcccc}
\toprule
\textbf{Method} & \textbf{V} & \textbf{A} & \textbf{D} & \textbf{L} \\
\midrule
Late Fusion*~\shortcite{Tang2017Multimodal}& 0.522 & 0.585 & 0.566 & 0.603 \\
Bi-LSTM*~\shortcite{zhao2021expression}              & 0.592 & 0.621 & 0.562 & 0.630 \\
TACOFormer*~\shortcite{li2023tacoformer}              & 0.603 & 0.635 & 0.578 & 0.662 \\
IANet*~\shortcite{LI2024IANet}              & 0.607 & 0.628 & 0.599 & 0.643 \\
\textbf{Ours}              & \textbf{0.642} & \textbf{0.695} & \textbf{0.601} & \textbf{0.689} \\
\bottomrule
\end{tabular}
\end{table}

\subsection{Ablation Study}

\begin{figure}[htbp]
\centering
\includegraphics[width=0.99\linewidth]{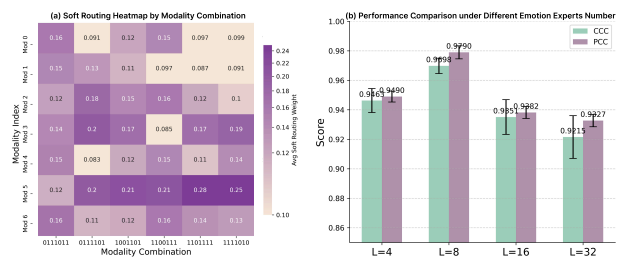}
\caption{(a) Average soft routing weights in the modality expert bank by modality combination. (b) Performance of the Emotion Expert Bank with varying numbers of experts.}
\label{fig:VAblation}
\end{figure}

% \begin{figure}[t]
%   \centering
%   \begin{subfigure}[b]{0.48\linewidth}
%     \centering
%     \includegraphics[width=\linewidth]{FIGURE/soft_routing_heatmap.pdf}
%     \caption{Average soft routing weights in the modality expert bank by modality combination.}
%     \label{fig:expert_heatmap}
%   \end{subfigure}
%   \hfill
%   \begin{subfigure}[b]{0.48\linewidth}
%     \centering
%     \includegraphics[width=\linewidth]{FIGURE/number_of_emotion_experts.pdf}
%     \caption{Performance of the Emotion Expert Bank with varying numbers of experts.}
%     \label{fig:expert_count_ablation}
%   \end{subfigure}
%   \caption{Analysis of Expert Bank Configurations}
%   \label{fig:Ablation}
% \end{figure}
To thoroughly evaluate the contribution of each proposed component and analyze key design choices within the proposed framework, we conduct comprehensive ablation studies. These experiments aim to demonstrate the necessity of each module for robust and fine-grained continuous emotion regression, and to provide insights into their specific functionalities. All ablation studies are performed on the DEAP dataset for regression, with additional results for classification provided in the Appendix.

\subsubsection{Contribution of Core Modules}
We first assess the individual impact of the framework's core modules by progressively removing them from the full model. Tab.~\ref{tab:ablation} reports the overall average performance of the full model and its ablated variants. As observed, removing any component results in notable performance declines, confirming their essential roles in robust and fine-grained emotion regression. Specifically, ablating the Emotion Expert Bank markedly reduces performance from 0.9698 to 0.9014, underscoring its crucial role in modeling fine-grained emotional prototypes. Removing the Cross-Modal Alignment Module also leads to a significant drop of 0.0126, highlighting its importance in addressing temporal and semantic misalignments for accurate emotion tracking. Omitting Soft Routing (CCC: 0.9698 vs. 0.9477) causes substantial degradation, emphasizing its necessity for dynamic resource allocation with diverse and incomplete inputs.

\begin{table}[htbp]
\centering
\caption{Ablation study on DEAP dataset: Overall average performance. (CCC / PCC $\uparrow$ MAE $\downarrow$)}
\label{tab:ablation}
\begin{tabular}{lccc}
\toprule
Module Setting & CCC & PCC & MAE \\
\midrule
w/o Emotion Expert Bank & 0.9014 & 0.9236 & 0.780 \\
w/o Cross-Modal Alignment & 0.9572 & 0.9614 & 0.443 \\
w/o Soft Routing & 0.9477 & 0.9491 & 0.512 \\
Full Model (Ours) & \textbf{0.9698} & \textbf{0.9790} & \textbf{0.321} \\
\bottomrule
\end{tabular}
\end{table}

\subsubsection{Modality Expert Bank's Adaptive Routing}
To gain deeper insight into the Modality Expert Bank's adaptive soft routing mechanism, we visualize the average routing weights based on input modality combinations. Fig.~\ref{fig:VAblation}(a) illustrates how routing weights dynamically adjust: when certain modalities are missing (indicated by `0' in the binary string on the X-axis), the model intelligently redistributes weights, often increasing reliance on available complementary modalities to compensate for the missing information. This visualization directly demonstrates how the soft routing enables flexible resource allocation and robust information integration, even when faced with arbitrary modality incompleteness. The varying weights across different experts and combinations confirm the dynamic and adaptive nature of the routing strategy, which is crucial for maintaining performance under diverse real-world conditions.

\subsubsection{Impact of Emotion Expert Count}
We also investigate the impact of the number of experts (L) within the Emotion Expert Bank on overall performance. As shown in Fig.~\ref{fig:VAblation}(b), we evaluate models with varying L values, which indicates that performance generally improves with an increasing number of experts up to a certain point, reflecting enhanced capacity to capture diverse emotional prototypes. However, beyond an optimal L, gains diminish or performance may even slightly degrade due to increased complexity or optimization challenges. These results justify the chosen L value for the Emotion Expert Bank, demonstrating a balance between model expressiveness and efficiency. Results under other missing modality rates are shown in the Appendix.

\section{Discussion}
Our hierarchical MoE framework demonstrates strong adaptability and robustness in real-world emotion recognition scenarios, achieving state-of-the-art results in continuous emotion regression on representative multimodal datasets like DEAP and DREAMER. Crucially, it maintains high prediction accuracy even under challenging conditions such as missing modalities and asynchronous sampling, directly addressing a key challenge in multimodal affective computing: reliable performance under imperfect, variable input conditions. This resilience, extending beyond controlled benchmarks to open-world settings, highlights its promise for truly deployable emotional intelligence systems and broadens the utility of MoE architectures in domains previously limited by assumptions of data reliability.

At the core of our framework is a novel dual-layer expert structure, designed for both expressiveness and adaptability. The Modality Expert Bank employs soft routing to dynamically weigh expert sub-modules, ensuring robust inference and flexible adaptation even with partial modality availability. Complementarily, the Emotion Expert Bank leverages a differential-attention routing strategy to flexibly attend to emotional prototypes, thereby capturing subtle individual or temporal affective nuances. This structural separation enhances the model’s ability to dynamically adapt to varying modality quality while preserving fine-grained emotional representation. Another key contribution is our explicit consideration of temporal misalignment across modalities, a challenge often overlooked. For instance, EEG signals and facial expressions frequently exhibit response lags~\cite{Aktürk2021Event-related, muukkonen2022representational}. To mitigate the defect, our model incorporates a cross-modal alignment module that learns to synchronize latent features both temporally and semantically, significantly improving fusion performance.

Despite the strong empirical results, the framework introduces additional computational overhead due to the use of multiple expert banks and routing operations. Moreover, the training and evaluation were conducted on datasets collected in controlled laboratory settings with limited diversity in user behavior and cultural background.

\section{Conclusion}
In this paper, we presented a novel hierarchical Mixture-of-Experts (MoE) framework for robust continuous emotion recognition under incomplete and asynchronous multimodal inputs, achieving state-of-the-art performance by effectively handling modality missingness and temporal inconsistencies. Future work will explore more lightweight architectures, online adaptation, user-aware learning, and validation on diverse in-the-wild datasets to enhance real-world applicability and reduce dependency on dense annotations.

\bibliography{aaai2026}


\newpage
\appendix
\section{Appendix}
\subsection{Modality-Specific Encoder}
Each modality sequence \( X_m \in \mathbb{R}^{T \times d_{in}} \) is processed by a dedicated encoder \( f_m(\cdot) \), producing feature representations:
\[ Z_m = f_m(X_m) \in \mathbb{R}^{T \times d} \]
We implement four types of encoders tailored to different modalities: EEG signals are encoded using EEGNet~\cite{Lawhern2018EEGNet} to capture spatiotemporal dependencies; facial expressions are processed via a structure-aware GCN that extracts frame-level features using EfficientNet-B0~\cite{tan2020efficientnetrethinkingmodelscaling}, followed by AU-specific embedding heads and a graph-based GCN module; dual-channel and single-channel physiological signals are encoded using shallow neural networks or lightweight CNN or MLP variants, depending on modality characteristics. To preserve modality-specific semantics, parameters are not shared across encoders. The resulting features \(\{Z_m\}\) are temporally aligned and serve as input to the expert routing and fusion modules.

\subsection{Dataset Pre-processing}
Our study utilizes four public emotion recognition datasets: DEAP, DREAMER, SEED, and AMIGOS. Each dataset presents unique characteristics in terms of collected modalities, experimental protocols, and annotation schemes. Below in table~\ref{tab:dataset_comparison}, we provide a detailed description and the specific preprocessing steps applied to each dataset.

% DEAP~\cite{Koelstra2012DEAP} is a widely-used dataset for emotion analysis. It consists of physiological and brain signals from 32 healthy participants (16 male, 16 female) while they watched 40 one-minute-long music videos. The dataset includes 32-channel EEG signals and peripheral physiological signals (PPS) such as galvanic skin response (GSR), blood volume pulse (BVP), and respiration. Participants self-annotated each video based on Valence, Arousal, Dominance, and Liking on a continuous scale from 1 to 9.

For the EEG signals in DEAP and Dreamer, we downsampled the original 512 Hz data to 128 Hz. We applied a bandpass filter from 4 Hz to 45 Hz to remove DC components and high-frequency noise. The data was segmented into 60-second trials, and a sliding window of 4 seconds with a 2-second overlap was used to create individual samples for our continuous prediction task. The PPS signals were also downsampled to 128 Hz and processed with a bandpass filter from 0.05 Hz to 2 Hz.

As for the 16-channel EEG in SEEDS, signals were downsampled from 1000 Hz to 200 Hz. We applied a bandpass filter from 1 Hz to 75 Hz. The data was segmented into 60-second trials. And we converted the discrete labels into a one-hot encoding format.

Moreover, for the AMIGOS datasets, we downsampled the original 128 Hz data to 64 Hz and applied a bandpass filter from 4 Hz to 45 Hz. The data was segmented into 4-second windows with a 2-second overlap. Similarly, the ECG data was downsampled and filtered (0.5-40 Hz). Video data was preprocessed to extract facial landmarks and other relevant features.

\begin{table*}[htbp]
    \centering
    \begin{tabular}{ccccc}
        \toprule
        \textbf{Features} & \textbf{DEAP} & \textbf{DREAMER} & \textbf{SEED} & \textbf{AMIGOS} \\
        \midrule
        \textbf{Num.} & 32 & 23 & 15 & 40 \\
        \textbf{Modality Types} & EEG, PPS & EEG, ECG, Video & EEG & EEG, ECG, Video \\
        \textbf{Stimuli} & Music Videos & Movie Clips & Movie Clips & Short/Long Videos \\
        \textbf{Trials} & 40 & 18 & 15 & 20 \\
        \textbf{Annotation} & Continuous (1-9) & Continuous (1-5) & Discrete (3 levels) & Continuous (1-9) \\
        \textbf{Dimensions} & V, A, D, L & V, A & V, A (inferred) & V, A, D \\
        \bottomrule
    \end{tabular}
    \caption{Comparison of DEAP, DREAMER, SEED, and AMIGOS datasets}
    \label{tab:dataset_comparison}
\end{table*}

\subsection{Performance on Complete Multimodal Data}

\begin{figure}[htbp]
\centering
\includegraphics[width=0.99\linewidth]{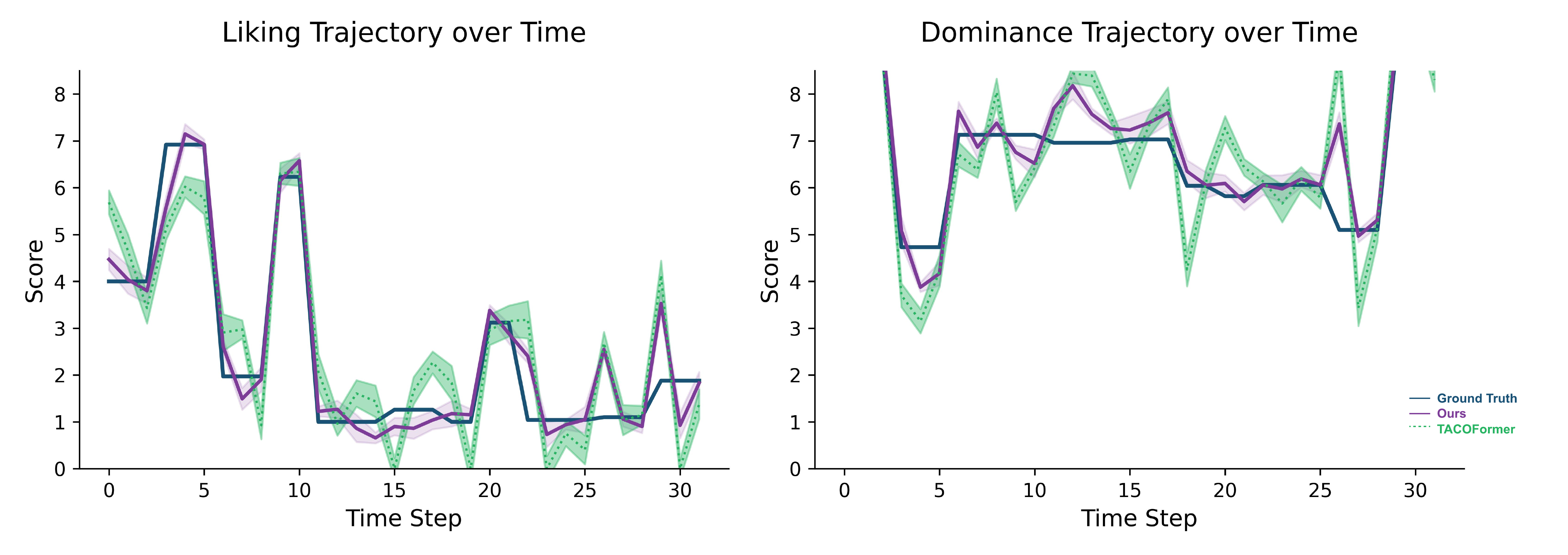}
\caption{Predicted trajectories compared with ground truth.}
\label{fig:D-L-trajectories-over-time}
\end{figure}

As shown in Fig.~\ref{fig:D-L-trajectories-over-time}, the predicted trajectories for both Dominance and Liking dimensions demonstrate our model's ability to effectively track the ground truth trends. Similar to the V-A dimensions, our model's predictions are notably smoother and more temporally consistent compared to the baseline methods, highlighting its robustness and capability to capture nuanced emotional changes across all annotated dimensions.

\subsection{Robustness under Modality Missing}
To provide a more comprehensive evaluation of our model's robustness to missing data, we extend the analysis from the main paper to a more constrained multimodal setting: the EEG+PPS and the EEG+EOG+EMG+EDA combination. This configuration is widely studied in emotion recognition and allows for a direct comparison with methods that rely solely on these two modalities. The experimental setup remains the same as described in the main paper, with the modality missing rate (r) ranging from 0.00 to 0.40.

As shown in Table~\ref{tab:modality_missing_EEGPPS} and~\ref{tab:modality_missing_EEGEOGEMGEDA}, our model consistently outperforms the baseline methods across all modality missing rates. These results further underscore the effectiveness of our hierarchical Mixture-of-Experts framework. The adaptive soft routing mechanism of our Modality Expert Bank successfully re-weights the available physiological signals, enabling robust information fusion even when one of the modalities is frequently unavailable. This demonstrates that our model's robustness is not limited to the all-modality configuration but generalizes well to more specific and constrained multimodal setups, reinforcing its suitability for real-world applications.

\begin{table*}[!h]
\centering
\begin{tabular}{lccccccccc}
\toprule
Missing Rate ($r$) & 0.00 & 0.05 & 0.10 & 0.15 & 0.20 & 0.25 & 0.30 & 0.35 & 0.4\\
\midrule
Late Fusion~\shortcite{Tang2017Multimodal}    & 0.8541 & 0.8112 & 0.7643 & 0.7444 & 0.6881 & 0.6534 & 0.5290 & - & -\\
TACOFormer*~\shortcite{li2023tacoformer}& 0.9144 & 0.8626 & 0.8312 & 0.8009 & 0.7446 & 0.6923 & 0.5656 & 0.5331 & 0.5104\\
\textbf{Ours (Full)} & \textbf{0.9522} & \textbf{0.9399} & \textbf{0.9205} & \textbf{0.9001} & \textbf{0.8824} & \textbf{0.8531} & \textbf{0.8496} & \textbf{0.8351} & \textbf{0.7677} \\
\bottomrule
\end{tabular}
\caption{CCC under EEG+PPS missing rates on the DEAP dataset (CCC $\uparrow$).}
\label{tab:modality_missing_EEGPPS}
\end{table*}

\begin{table*}[!h]
\centering
\begin{tabular}{lccccccccc}
\toprule
Missing Rate ($r$) & 0.00 & 0.05 & 0.10 & 0.15 & 0.20 & 0.25 & 0.30 & 0.35 & 0.4\\
\midrule
IANet*~\shortcite{LI2024IANet}& 0.9002 & 0.8434 & 0.8399 & 0.8217 & 0.7363 & 0.7192 & 0.5564 & 0.5499 & 0.5665\\
\textbf{Ours (Full)} & \textbf{0.9131} & \textbf{0.9055} & \textbf{0.8998} & \textbf{0.8764} & \textbf{0.8616} & \textbf{0.8231} & \textbf{0.8005} & \textbf{0.7872} & \textbf{0.7364} \\
\bottomrule
\end{tabular}
\caption{CCC under EEG+EOG+EMG+EDA missing rates on the DEAP dataset (CCC $\uparrow$).}
\label{tab:modality_missing_EEGEOGEMGEDA}
\end{table*}

\begin{figure}[htbp]
    \centering
    \includegraphics[width=1.0\linewidth]{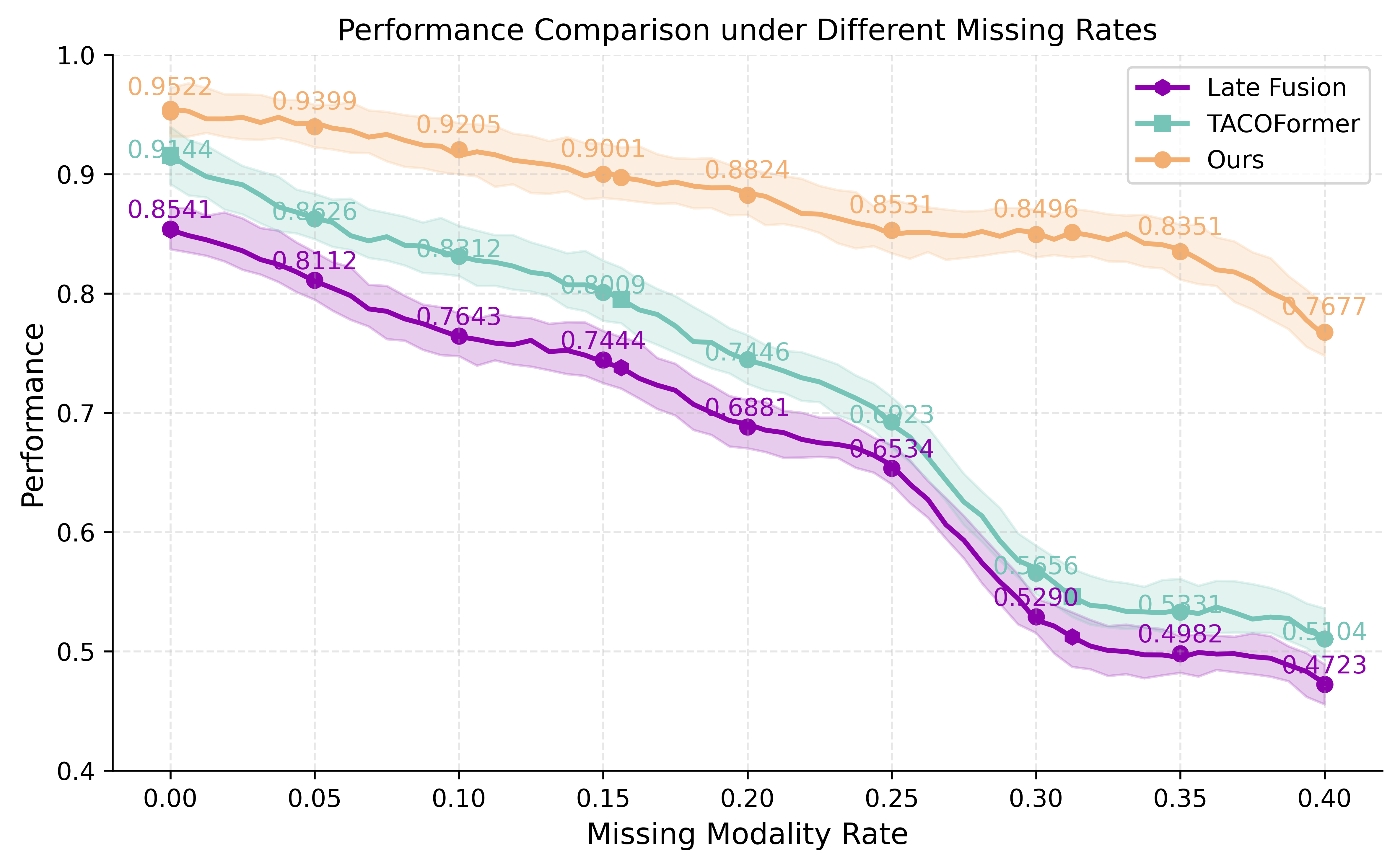}
    \caption{Performance comparison under varying modality missing rates with the configuration of the EEG and the PPS}
    \label{fig:missing_trends_eeg_pps}
\end{figure}

\begin{figure}[htbp]
    \centering
    \includegraphics[width=1.0\linewidth]{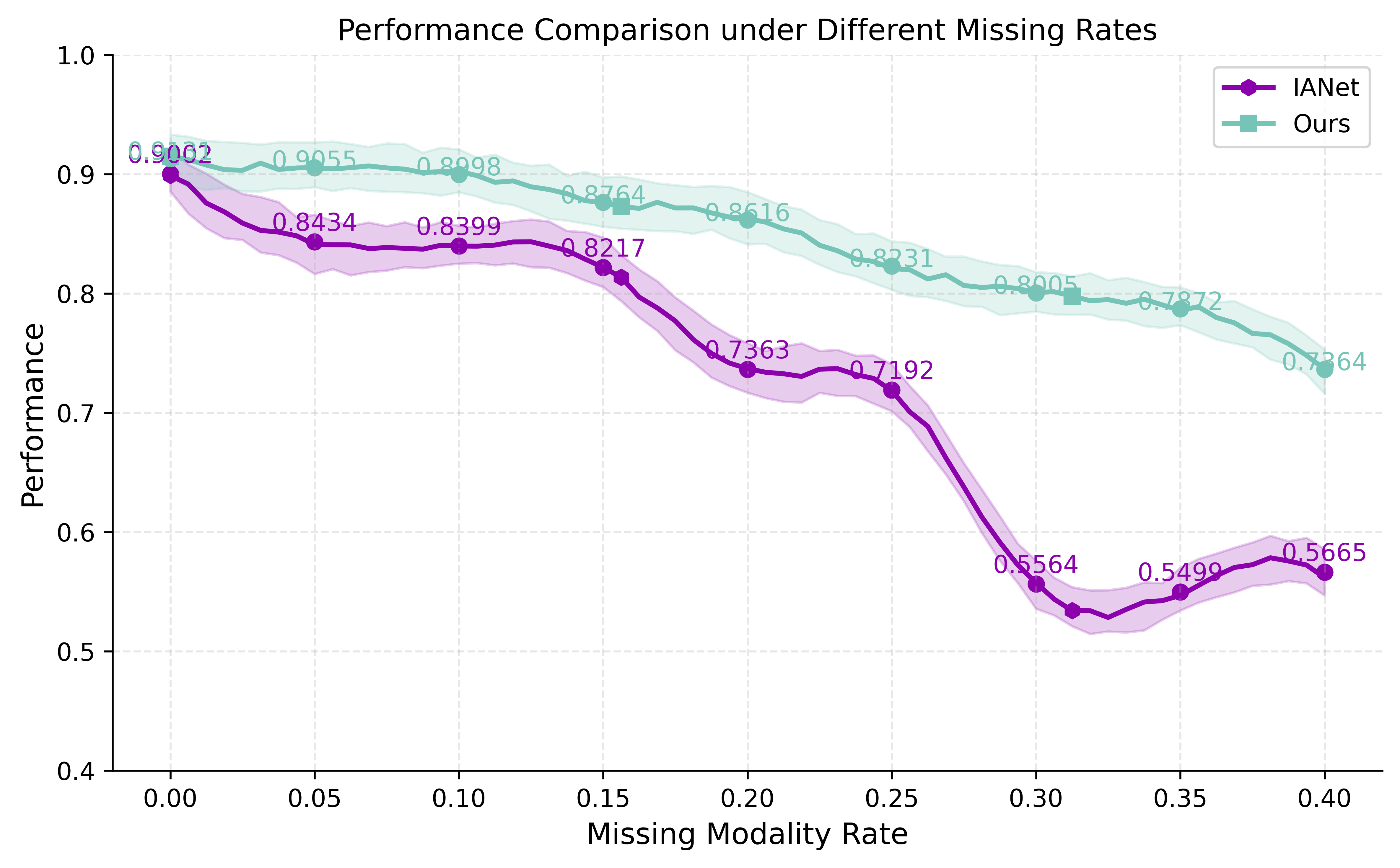}
    \caption{Performance comparison under varying modality missing rates with the configuration of the EEG, EOG, EMG, and the EDA.}
    \label{fig:missing_trends_eeg_eog_emg_eda}
\end{figure}

To supplement the tables and analysis provided in the main paper, we visualize the performance trends of our model and baselines under various modality missing rates. As shown in Figure~\ref{fig:missing_trends_eeg_pps} and Figure~\ref{fig:missing_trends_eeg_eog_emg_eda}, our model demonstrates a more robust and graceful degradation curve compared to the baselines. This visual evidence further confirms that our framework is more resilient to modality absence and maintains superior performance even as the missing rate increases.

\subsection{Discrete Classification with Adaptive Modality Handling}
To comprehensively evaluate the model's performance, we conducted further experiments on binary classification tasks using the DEAP dataset. This section supplements the results on different physiological signal subsets, such as high/low Arousal and high/low Valence. These findings complement the regression task results presented in the main paper, collectively demonstrating our model's effectiveness and generalization capability across various emotion recognition tasks. Detailed performance metrics, including Accuracy and F1-Score, will be presented in the following tables~\ref{tab:binary_results_subsets}.

\begin{table*}[!h]
\centering
\caption{Performance on the EEG and Facial (V-A).}
\label{tab:binary_results_subsets}
\begin{tabular}{lccccc}
\toprule
\multirow{2}{*}{\textbf{Method}} &\multirow{2}{*}{\textbf{Modalities}} & \multicolumn{2}{c}{\textbf{Valence}} & \multicolumn{2}{c}{\textbf{Arousal}} \\
\cmidrule(lr){3-4} \cmidrule(lr){5-6}
 & &  ACC & F1 & ACC & F1 \\
\midrule
TACOformer~\shortcite{li2023tacoformer} & EEG, FACIAL  & 91.59 & - & 92.02 & - \\
CAFNet   & EEG, FACIAL & 94.99 & 95.40 & 95.89 & 96.09 \\
\textbf{Ours}  & EEG, FACIAL & \textbf{98.15} & \textbf{98.77} & \textbf{98.63} & \textbf{98.91}\\
\midrule
Husformer &EEG, EOG, EMG, EDA& 90.67 & 90.74 & 91.33 & 91.35 \\
IANet~\shortcite{LI2024IANet} &EEG, EOG, EMG, EDA& 97.42 & 97.41 & 97.56 & 97.55 \\
\textbf{Ours}  &EEG, EOG, EMG, EDA&  \textbf{97.57} & \textbf{97.78} & \textbf{98.24} & \textbf{98.31}\\
\bottomrule
\end{tabular}
\end{table*}

\end{document}